# New phases in the Ba–Ce(M)–O systems (M = Ga, In)

N.I. Matskevich (*Nikolaev Institute of Inorganic Chemistry, Siberian Branch of the Russian Academy of Science, Novosibirsk, Russia),* T.I. Chupahina, G.V. Bazuev *(Institute of Solid State Chemistry, Ural Branch of the Russian Academy of Science, Ekaterinburg, Russia),* F.A. Kuznetsov (*Nikolaev Institute of Inorganic Chemistry, Siberian Branch of the Russian Academy of Science, Novosibirsk, Russia)*

E–mail: nata@che.nsk.su

**Abstract**

A new and known barium cerates doped by In and Ga were synthesized in air at 1700 K. The synthesis was performed by solid state reactions from $BaCO_3$, $CeO_2$, $In_2O_3$ or $Ga_2O_3$. X–ray diffraction profiles of the powdered samples were obtained with DRON–UM–1 and STADI–P diffractometers using monochromatized $CuK_\alpha$ radiation. The following compositions of new phases were obtained: $BaCe_{1-x}Ga_xO_{3-\delta}$ (x = 0.05; 0.1) and $BaCe_{1-x}In_xO_{3-\delta}$ (x = 0.25, 0.5). They crystallize in the orthorhombic space group *Pnma.* The lattice parameters were calculated by full profile Rietveld method. Based on analysis of literature data it was possible to assume that increasing of tolerance factors from $BaCeO_3$ up to $BaCe_{1-x}Ga_xO_{3-\delta}$ ($BaCe_{1-x}In_xO_{3-\delta}$) led to the increasing of stability and transport properties of Ga and In doped barium cerates.

## 1. Introduction

Complex cerium oxides with perovskite related structures are promising ceramic materials for application in hydrogen sensors, electrocatalytic reactors for hydrogen separation, and also as electrolytes in intermediate temperature fuel cells operating on hydrocarbon fuels. The substituted solid solution materials based on the perovskite oxide $BaCeO_3$, in which M replaces some percents of Ce, are typical examples of proton conducting oxides of this class. In these compounds M is usually rare earth element such as Y, Gd, Nd, Yb etc. The compound of $BaCe_{0.9}Nd_{0.1}O_{3-\delta}$ show the highest conductivity among Ba–Ce(RE)–O substituted solid solutions (5.5 x $10^{-1}$ S/cm at 1000 C) [1]. The systematic study of perovskite type oxides, which have proton conductivity, started from 1980 [2]. Several classes of materials such as doped zirconates, doped cerates, doped niobates etc. were studied during ten-twenty years [1-13]. Paper [13] devoted to the investigation of protonic conduction in $Ba_2In_2O_5$ where transport properties of this compound were studied by conductivity and EMF methods as a function of temperature and oxygen activity.

As well known, in the next generation of energy devices there is considerable emphasis on moving to lower temperature of operation. One of the important problems in this direction will be addressed to find suitable materials that operate well at lower temperatures. Good progress can be achieved using perovskites such as doped $BaCeO_3$ or perovskite-related materials such as $Ba_3(Ca_{1+x}Nb_{2-x})O_{9-\delta}$. For this reason, $BaCeO_3$ phases doped by different metals seem to be perspective materials for future application.

Basing on the fact that both $BaCeO_3$ and $Ba_2In_2O_5$ phases have good perspective as proton conductors, we decided to prepare new unknown compounds in the $BaO$–$CeO_2$–$In_2O_3$ and $BaO$–$CeO_2$–$Ga_2O_3$ systems.

In this study we focus our attention on the synthesis of $BaIn_xCe_{1-x}O_{3-\delta}$ and $BaGa_xCe_{1-x}O_{3-\delta}$. There is no information about compound of $BaGa_xCe_{1-x}O_{3-\delta}$. There is information only about compound of $BaIn_xCe_{1-x}O_3$ with x up to 0.25.

## 2. Experimental

### *2.1. Phases in the Ba–Ce–O system doped by Ga*

Polycrystalline samples of $BaCe_{1-x}Ga_xO_{3-\delta}$ (x = 0.05; 0.1; 0.2; 0.5 ) were prepared by solid state reaction using starting materials $BaCO_3$, $CeO_2$, and $Ga_2O_3$. These starting materials were weighed in the corrected ratios, mixed well and milled in planetary mill with agate balls during several hours with intermediate regrinding. Then powers were pressed into pellets (P = 3000 kg · $sm^{-2}$) and calcinated at 1100 K, 1300 K, 1400 K, 1700 K with intermediate regrinding. These conditions of synthesis were resulted in phase pure $BaCe_{1-x}Ga_xO_{3-\delta}$ samples at x=0.05 and x=0.1. The samples at x=0.2 and 0.5 contents $Ga_2O_3$ as impurity phase.

### *2.2. Phases in the Ba–Ce–O system doped by In*

In paper [13] the synthesis, structure and electrochemical properties of In-doped $BaCeO_3$ ($BaCe_{1-x}In_xO_{3-\delta}$) is described up to x = 0.25. Authors mention that sample with x = 0.25 is not phasing pure $BaCe_{0.75}In_{0.25}O_{3-\delta}$. There are impurities of $BaCeO_3$ and $CeO_2$ in this sample. There are no parameters of lattice of $BaCe_{1-x}In_xO_{3-\delta}$ in paper [13].

We prepared $BaCe_{1-x}In_xO_{3-\delta}$ phases with x = 0.05, 0.1, 0.25 and 0.5. It was found that $BaCe_{1-x}In_xO_{3-\delta}$ could be obtained as a single products up to x = 0.5.

The samples were prepared starting with analytical reagent grade $BaCO_3$, $CeO_2$, and $In_2O_3$. The regimes of synthesis were the same like for $BaCe_{1-x}Ga_xO_{3-\delta}$ phases.

## 3. Results and discussion

X-ray diffraction profiles of powdered samples were obtained with DRON–UM–1 and STADI–P diffractometers using monochromatized $CuK_\alpha$ radiation. The following compositions of new phases were obtained: $BaCe_{1-x}In_xO_{3-\delta}$ (x = 0.25, 0.5) and $BaCe_{1-x}Ga_xO_{3-\delta}$ (x = 0.05; 0.1). They crystallize in the orthorhombic space group *Pnma.* The lattice parameters were calculated by full profile Rietveld method.

Intensity data for $BaCe_{1-x}M_xO_{3-\delta}$ were collected over a $2\Theta$ range of 10–70°. X-ray data for $BaCe_{0.9}Ga_{0.1}O_{3-\delta}$ are presented at Fig.1.

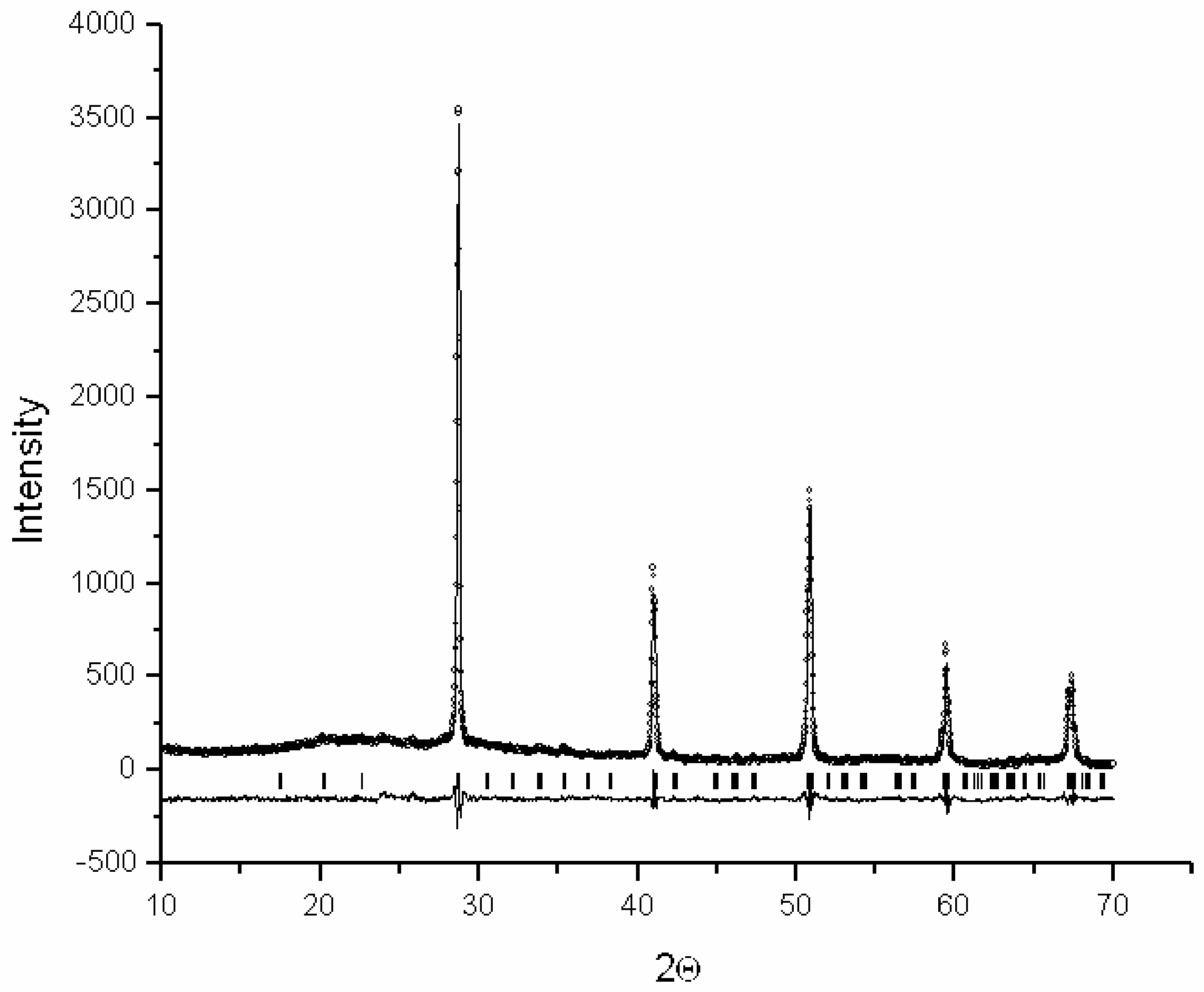

FIG. 1. X-ray diffraction for $BaCe_{0.9}Ga_{0.1}O_{3-\delta}$

The X-ray diffraction analysis indicated that the solid state reaction

$$BaCO_3 + (1–x)CeO_2 + (x/2)In_2O_3 = BaCe_{1–x}In_xO_{3–\delta} + yO_2 + CO_2$$

is finished under the applied calcinations conditions (1700 K).

The same X-ray technique like for $BaCe_{1–x}Ga_xO_{3–\delta}$ was used to obtain X-ray diffraction pattern for In-doped $BaCeO_3$.

Below (Fig. 2) we present X-ray diffraction pattern for $BaCe_{0.75}In_{0.25}O_{3–\delta}$ and $BaCe_{0.5}In_{0.5}O_{3–\delta}$. As it is possible to see, both samples are phase pure without impurities of $BaCeO_3$ and $CeO_2$.

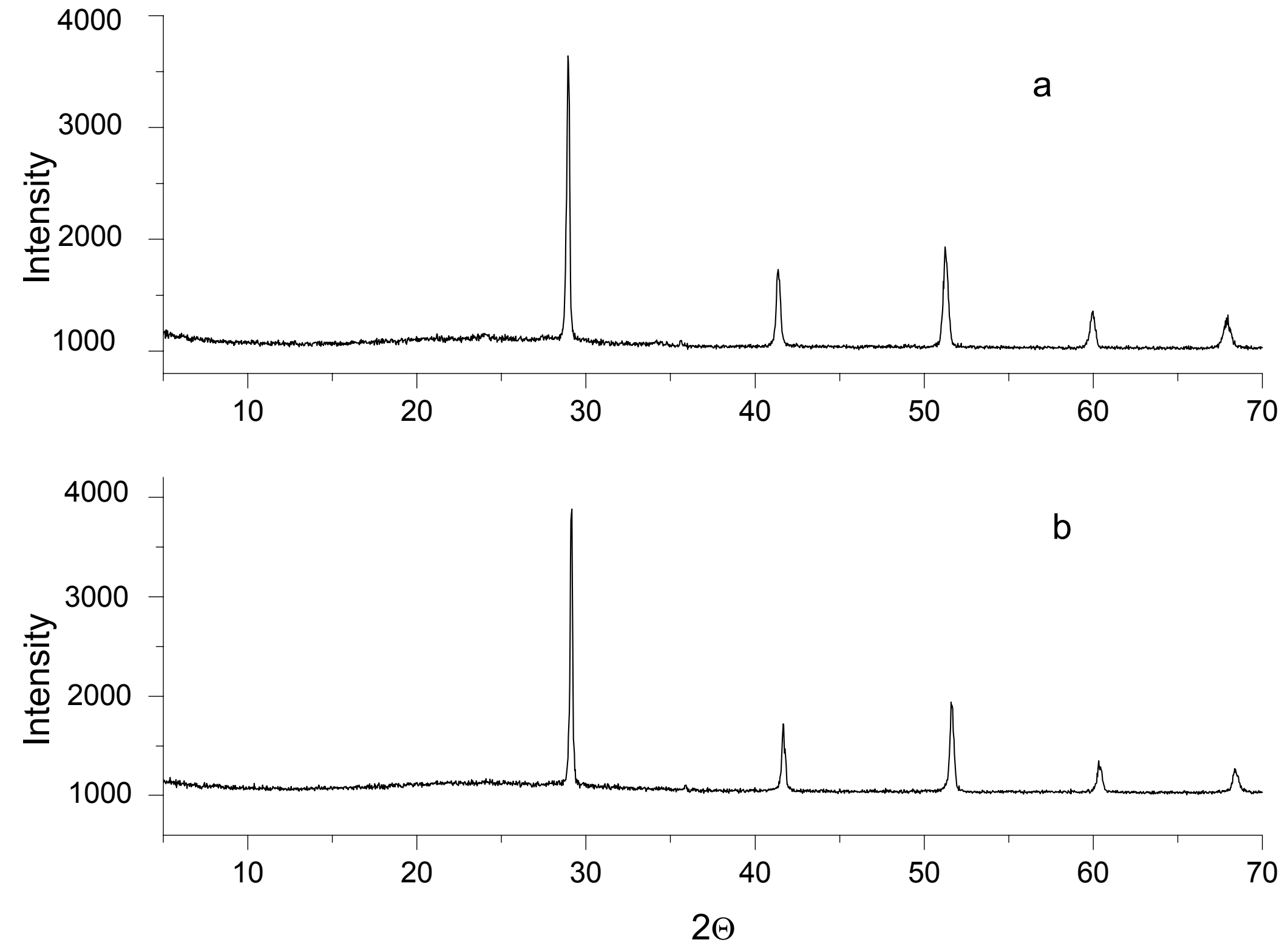

FIG. 2. X-ray diffraction pattern for $BaCe_{0.75}In_{0.25}O_{3–\delta}$ (a) and $BaCe_{0.5}In_{0.5}O_{3–\delta}$ (b)

Lattice parameters for $BaGa_xCe_{1–x}O_{3–\delta}$ and $BaIn_xCe_{1–x}O_{3–\delta}$ are reported in Table 1.

Table 1

| Compounds | a, nm | b, nm | c, nm |
|---|---|---|---|
| $BaGa_{0.1}Ce_{0.9}O_{3-\delta}$ | 0.623413 (31) | 0.621236 (27) | 0.877180 (46) |
| $BaIn_{0.1}Ce_{0.9}O_{3-\delta}$ | 0.620943 (78) | 0.619953 (54) | 0.876677 (53) |
| $BaIn_{0.25}Ce_{0.75}O_{3-\delta}$ | 0.616962 (42) | 0.617121 (85) | 0.872161 (73) |
| $BaIn_{0.5}Ce_{0.5}O_{3-\delta}$ | 0.612978 (54) | 0.612032 (21) | 0.866364 (44) |

As it is possible to see from Table 1, lattice parameters for $BaIn_xCe_{1-x}O_{3-\delta}$ are decreased with increasing of In-content. This decrease of parameters reflects the substitution of small cations $In^{3+}$ (r=0.080 nm) for $Ce^{4+}$ (r=0.087 nm) and decreasing of oxygen content due to decreasing of valence of B-cation and formation of additional oxygen vacancies.

As it has been mentioned before, doped cerates of alkali-earth metals are perspective materials for fuel cells, electrocatalytic reactors, etc. For successful application of new materials it is necessary to know such important characteristics as stability and transport properties. From this point of view it would be interesting to predict a direction in which stability and transport properties would be changed from pure $BaCeO_3$ up to doped $BaCeO_3$. The stability of $BaCeO_3$ -based materials not only depends on thermal conditions but also on the nature of B-site dopant ion. It has been suggested in paper [1] that stability increases when the dopant ion is small and does not exhibit strong base properties.

In the literature there are no data on stability of $BaIn_xCe_{1-x}O_{3-\delta}$ and $BaGa_xCe_{1-x}O_{3-\delta}$. We will try to predict stability of these compounds based on consideration of correlations between tolerance factors and stabilities of $BaCeO_3$ and $SrCeO_3$. We used values of ionic radius taken from paper [14] to calculate tolerance factors (see, Table 2).

Table 2

| Cation | Radius (nm) | Cation | Radius (nm) |
|---|---|---|---|
| O2– | 0.121 | Y3 + | 0.089 |
| Ba2 + | 0.134 | Gd3 + | 0.094 |
| Sr2 + | 0.112 | In3 + | 0.079 |
| Ce4 + | 0.080 | Ga3 + | 0.047 |

On the basis of these data we calculated tolerance factors according to formula:
$t = (r_A + r_O)\ [\sqrt{2}\ (r_B + r_O)]^{-1}$. The data are resulted in Table 3.

Table 3

| Compound | t | Compound | T |
|---|---|---|---|
| $BaCeO_3$ | 0.8970 | $BaCe_{0.9}Gd_{0.1}O_{3-\delta}$ | 0.8908 |
| $SrCeO_3$ | 0.8196 | $BaCe_{0.9}In_{0.1}O_{3-\delta}$ | 0.8975 |
| $BaCe_{0.9}Y_{0.1}O_{3-\delta}$ | 0.8931 | $BaCe_{0.9}Ga_{0.1}O_{3-\delta}$ | 0.9120 |

Based on the standard molar enthalpies of formation of $SrCeO_3$ and $BaCeO_3$ measured in paper [15] and enthalpies of formation of SrO, BaO, $CeO_2$ taken from reference book [16] we calculated formation enthalpies for $SrCeO_3$ and $BaCeO_3$ from oxides as following:

$SrO + CeO_2 = SrCeO_3$ $\Delta_{ox}H^0$ (298.15 K) = –5.7 ± 2.7 kJ/mol

$BaO + CeO_2 = BaCeO_3$ $\Delta_{ox}H^0$ (298.15 K) = –51.6 ± 2.5 kJ/mol

This value for barium cerate is comparable to value of $\Delta_{ox}H^0$ for barium cuprate (see, for example, paper [17]).

As it is possible to see, the increase of tolerance factor from $SrCeO_3$ up to $BaCeO_3$ correlates with increasing of the stability. On this basis it is possible to assume, that barium cerates doped by In and Ga with higher tolerance factor will be more stable than pure $BaCeO_3$. It confirmed the assumption of paper [1].

As it was reported in paper [1], transport properties of barium cerates were higher than strontium cerates. The higher transport properties of $BaCeO_3$ correlate with higher enthalpy of hydration. Hydration enthalpies of $BaCeO_3$ and $SrCeO_3$ are –172 kJ/mol [18] and –157 kJ/mol [19] correspondingly. The increase of tolerance factor from $SrCeO_3$ up to $BaCeO_3$ correlates with increasing of hydration enthalpy. On this basis it is possible to assume, that barium cerates doped by In and Ga with higher tolerance factor will have higher transport properties than $SrCeO_3$.

## 4. Conclusion

We report synthesis of new phases in the Ba–Ce (M)–O systems (M = In, Ga). The following compositions of new phases were obtained: $BaCe_{1-x}Ga_xO_{3-\delta}$ (x = 0.05; 0.1) and $BaCe_{1-x}In_xO_{3-\delta}$ (x = 0.25, 0.5). We compared tolerance factors, stability, hydration enthalpies and transport properties for strontium and barium cerates. We assumed that tends of changing of stability and transport properties from $SrCeO_3$ up to $BaCeO_3$ are the same as from pure $BaCeO_3$ up to doped $BaCeO_3$. On this basis it was possible to assume that $BaIn_xCe_{1-x}O_{3-\delta}$ and $BaGa_xCe_{1-x}O_{3-\delta}$ would be more stable and have higher transport properties than pure $BaCeO_3$.

*This work is supported by Special program of interdisciplinary projects performed by scientists from Siberian Branch and Ural Branch (project Nr. 202).*